\documentstyle[twoside,fleqn,espcrc2,epsf]{article}

\newcommand{\bbg}{b\bar{b}g}

\title{Coarse lattice results for glueballs and hybrids.}

\author{Mike Peardon\address{Department of Physics and Astronomy, 
                             University of Kentucky,
                             Lexington, KY 40502-0055, USA.}
        }

\begin{document}
\pagestyle{empty}

\begin{abstract}
A review of new results from lattice simulations of glueballs and 
heavy-quark hybrid mesons is presented. 

\end{abstract}

\maketitle

\section{INTRODUCTION}
Glueballs and hybrid mesons are both colour-singlet bound states containing
excitations of gluonic degrees of freedom and so are absent in the quark model,
but expected in QCD. 
These gluonic excitations are a highly non-perturbative aspect of the theory 
and so 
making reliable predictions requires robust methods.
Unambiguous experimental identification of these states has proven 
to be difficult. At present, the two lightest glueballs, the scalar and tensor,
have candidate resonances; the $f_0(1500)$ \cite{Anis94} and $f_J(1710)$ 
(scalar) and the $\xi(2230)$ \cite{Balt85,Goda97} (tensor). 
It now seems likely that the scalar glueball is mixed strongly with near-by 
conventional mesons. 
Recently, new data for the light-quark hybrid candidate, with exotic quantum 
numbers $1^{-+}$ have been reported \cite{BNL97}, close to the latest lattice
predictions \cite{UKQCD96,MILC97} but,
to date, no signal for a heavy-quark hybrid has been 
reported. More details on the current status of phenomenological 
interpretations of experimental resonances are presented in Ref. \cite{Clos97}. 

Clearly then, reliable lattice data for the properties of these bound states
are helpful, both to interpret and predict experiment. Information beyond the 
masses of these states would seem to be required; the clearest example is 
the scalar glueball candidates, $f_0(1500)$ and 
$f_J(1710)$, which are both consistent with lattice mass predictions in the 
pure-gauge theory \cite{UKQCD93,Sext95,Morn97a} of 1600 MeV with 
systematic errors of approximately 100 MeV.

Simulations of hybrids with light quarks were reported to the conference
\cite{UKQCD96,MILC97}. 
In discussing hybrids, this review will concentrate on developments in the
study of the heavy-quark system, which is a natural
starting point for theoretical studies of hybrids, since the Born-Oppenheimer 
approximation can be used to separate the quark and gluon degrees of freedom. 

The presence of gluonic excitations in the rather heavy gluonic
states makes them a challenging problem for lattice simulations. The vacuum 
fluctuations in the constituent gluon fields tend to be rather large and so the 
rapidly-decaying Euclidean correlators which must be measured in a Monte-Carlo 
calculation become dominated by statistical noise for rather small source-sink 
time separations and extracting the masses becomes difficult. Studying
other properties is an even greater numerical problem. This year, 
simulations on improved, anisotropic lattices (with spatial lattice spacing, 
$a_s \ll a_t$, the temporal lattice spacing) were performed,
with encouraging results. This technique shows promise as an efficient method 
for simulation of gluonic bound states as it tries to incorporate the cost 
advantages of coarse lattices without reducing the energy
resolution too far. 

In this review, new results for glueball simulations and the gluonic
excitations of the static inter-quark potential from anisotropic lattices, as
well as NRQCD studies on large ensembles of Wilson action configurations will be
summarised. The direction of present and future work will be discussed.

\section{ANISOTROPIC LATTICES}

  In any attempt to simulate a bound state efficiently on a coarse lattice, it 
is important to recognise the two scales in the problem: 
the state's size and its mass. Glueballs have been estimated to have a size of 
about 0.5 to 0.8 fm (although some studies suggest the scalar glueball might 
be significantly smaller), while their masses are more than 1.5 GeV. For the
wave function to be distributed over three grid points, a spatial lattice
spacing of about 0.2 fm is required, while setting the temporal lattice spacing 
by demanding $m_G a_t<1$, which allows the correlator to be resolved over
sufficient time steps from accessible statistics, needs $a_t<0.1$ fm. 

  The construction of anisotropic lattice actions using mean-field improved
perturbation theory is dealt with in Refs. \cite{Morn96,Alfo9x} and (except
where specifically noted) all anisotropic simulation results presented here
are from this action. This action was designed to be $O(a_s^2)$ improved and to
include terms that directly couple links only on adjacent time slices to ensure
the free gluon propagator has no spurious modes. In all
cases considered, the effective masses computed from correlators between 
identical source and sink operators converged to their plateaus from above, 
illustrating positivity of the transfer matrix for this action.

  Using mean-link improved perturbation theory to determine the coefficients in
the action has been shown to reduce the renormalisation of the bare anisotropy 
to the level of a few percent. The true anisotropy
can be measured by computing the static potential along distinct lattice axes.
  
\section{GLUEBALLS}
\begin{figure}[t]
\leavevmode
\epsfxsize=2.9in\epsfbox{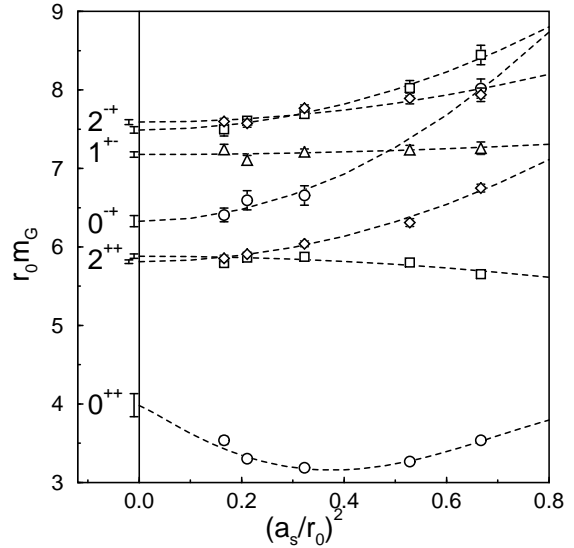}
\caption{Scaling behaviour of the lightest five glueball states from
5:1 anisotropic lattice simulations. Data from 4
irreps are included, with symbols $\circ=A_1, \Box=E, \triangle=T_1,
\diamond=T_2$. 
\label{fig:scaling}}
\end{figure}

  The scaling properties and continuum spectrum of glueballs has been studied 
extensively using anisotropic
lattices \cite{Morn97a,Morn97b}. The cut-off dependence of the lightest five 
glueball
states, which have $J^{PC}$ quantum numbers $0^{++},2^{++},0^{-+},1^{+-}$ and
$2^{-+}$ is shown in Fig.~\ref{fig:scaling}. Data from five simulations at
an input anisotropy of 5:1 are included on the plot and, for all but the scalar
glueball, fits to the anticipated leading discretisation error for this range 
of lattice spacings, 
\begin{equation}
\varphi(x) = r_0 m_G + c_1 (a_s/r_0)^4 
\label{eqn:phi4fitfunc}
\end{equation}
are included. For the scalar, a four parameter fit which models an $\alpha_s
a_s^2$ along with the $a_s^4$ term is used. 
\begin{figure}[t]
\leavevmode
\epsfxsize=2.9in\epsfbox{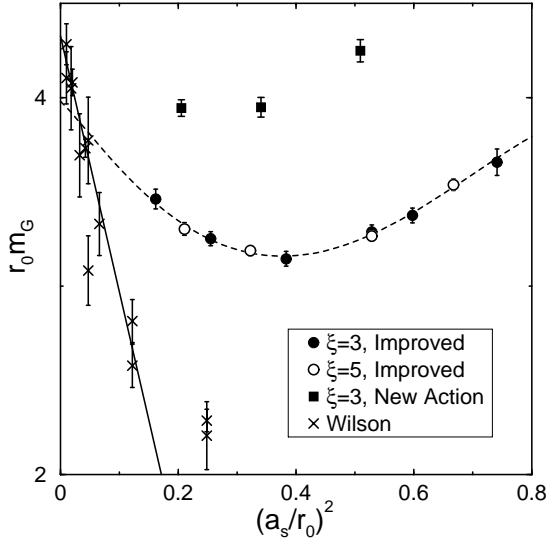}
\caption{The scaling behaviour of the $0^{++}$ glueball for three different
actions. \label{fig:scalar}}
\end{figure}
\begin{figure}[t]
\leavevmode
\epsfxsize=2.9in\epsfbox{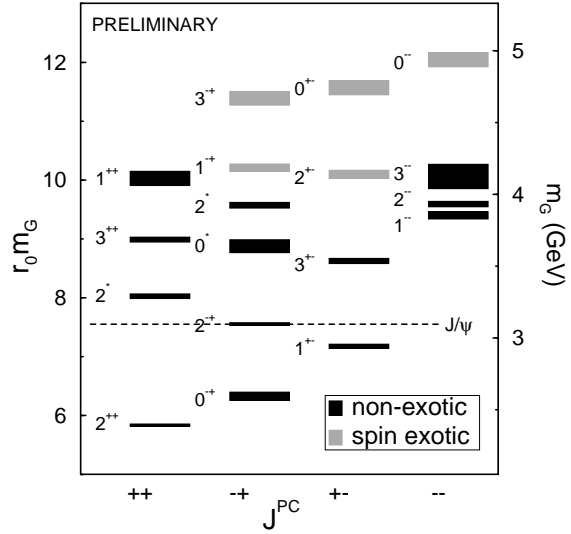}
\caption{The continuum limit for the higher levels in the $SU(3)$ glueball 
spectrum \label{spectrum}. Data are from 5 simulations on 5:1 anisotropic 
lattices \protect{\cite{Morn97b}}}
\end{figure}

  The scalar glueball shows significant scaling violations even at 0.2 fm. Fig.
\ref{fig:scalar} shows in more detail the scaling properties of the scalar
for both the Wilson and anisotropic, improved action (as well as a new
choice of anisotropic lattice action, which will be discussed later). 
At a lattice spacing, $a_s\approx 0.3$ fm, the dimensionless mass, 
$r_0 m(0^{++})$ is about $25-30\%$ lower than continuum estimates. The scalar
glueball is believed to be a small state, and might never be well
simulated on a coarse lattice with a Symanzik-improved action. The picture is 
complicated somewhat by the
existence of a critical end-point in the space of fundamental-adjoint 
couplings \cite{FundAdj} at which the correlation length in the scalar channel 
is seen to diverge. If this is the cause of
the poor scaling of the lightest glueball, then it might be possible to
construct actions whose renormalisation trajectory steers clear of the
problematic fixed point neighbourhood. A class of actions of this form is under 
investigation with some promising initial findings. Included in 
Fig.~\ref{fig:scalar} are three simulations performed with such an action. These
points are much closer to the continuum predictions and show only a few
percent scaling violations. Other glueballs are unaffected.
For SU(2), a similar, but less severe dip is found \cite{Trot} and the authors 
argue that the cut-off effects are reduced significantly with the use of the 
Landau gauge definition of the mean-link parameters, $u_\mu = \langle 
\frac{1}{3} \mbox{Re Tr }U_\mu\rangle_{\rm Landau}$ rather than the plaquette
definition used in {\it e.g.} Ref.~\cite{Morn97a}. 
The reduced severity of the dip here
is still consistent with the fixed point idea as the point is further
from the fundamental axis in SU(2) so should have a weaker influence. 
It is,
however, important to note that these points are not necessarily mutually 
exclusive; the cause of the dip could be due in part to all of these effects. 
The discussion should be resolved with more reliable information on the
size of the scalar, as well as simulations of SU(3) glueballs
with the Landau-defined mean-link parameter and studies of new actions.
It would also be interesting to see if the use of fixed-point
techniques \cite{Hase97} can determine a better anisotropic action. 
\begin{table}[thb]
\setlength{\tabcolsep}{5mm}
\begin{center}
\caption{Continuum higher glueball masses in pure-gauge SU(3) (Preliminary). 
$r_0^{-1}=410(20)$ MeV is used to set the scale. \label{tab:glueballs}}
\begin{tabular}{lcc}
\hline
 $J^{PC}$ & $r_0m_G$ & Mass (MeV) \\
\hline
$2^{++}$ & $5.84(2)$ & $2390\pm120$ \\
$0^{-+}$ & $6.33(7)$ & $2590\pm130$ \\
$1^{+-}$ & $7.18(4)$ & $2940\pm140$ \\
$2^{-+}$ & $7.49(4)$ & $3070\pm150$ \\
$2^{*++}$& $8.02(4)$ & $3290\pm160$ \\
$3^{+-}$ & $8.63(5)$ & $3540\pm170$ \\
$0^{*-+}$& $8.9(1)$  & $3640\pm180$ \\
$3^{++}$ & $9.0(1)$  & $3690\pm180$ \\
$1^{--}$ & $9.4(1)$  & $3850\pm190$ \\
\hline
\multicolumn{3}{c}{lightest spin-exotic glueball} \\
$2^{+-}$ & $10.1(1)$ & $4100\pm200$ \\
\hline
\end{tabular}
\end{center}
\end{table}

Fig.~\ref{spectrum} shows the continuum limit of the spectrum of higher 
glueball 
states from five $\xi=5$ anisotropic simulations. For all these states, the 
simple function of Eqn. \ref{eqn:phi4fitfunc} fits the data well and so was 
used in determining the continuum limits. No uncertainty from the scale 
is included in Fig. \ref{spectrum}, however an estimate of
this systematic effect is included in Table \ref{tab:glueballs}. 
The data show that 
that, for the pure-gauge theory, the pseudoscalar glueball is heavier 
than the tensor.  This spectrum is still a preliminary result as ambiguities
in the interpretation of the data remain: the assignment of a continuum spin
label requires the resolution of degeneracies in the continuum limit across 
lattice irreps and the lattice operators creating more 
massive states can also overlap with both two light glueballs and torelon pairs 
(flux-tube-like excitations winding around the periodic boundaries of the
finite lattice). The lightest spin-exotic state in this preliminary analysis
is found to be very
massive, with $J^{PC}=2^{+-}$, $m_G=4.1(2)$ GeV. These results are in 
agreement with the existing Wilson action data \cite{UKQCD93} in this energy 
range. 

  In a lattice Hamiltonian formulation, the dynamics in the temporal direction 
are 
left continuous, and masses can be extracted by computing the energy
eigenstates. The Symanzik improvement scheme can also be applied to choose a 
spatial discretisation with reduced contaminations from the finite cut-off.
This idea is discussed in Ref. \cite{Luo97} and Hamiltonian methods have been 
applied to computations of the glueball spectrum \cite{Luo96}.

  Anisotropic lattice technology allows more properties of glueballs to be
explored. A new calculation \cite{Nenk97} to estimate the production rate of 
glueballs in the radiative decay of charmonium
($J/\psi\rightarrow\gamma G$) is a good example of this. Here, the matrix
elements appropriate for creation of the lightest three glueball states (the
scalar, tensor and pseudoscalar) of the form 
$\langle G |\;\mbox{Tr } F F \;| 0 \rangle$ are 
under investigation. 

  The SESAM collaboration \cite{Bali97} has investigated the scalar and tensor 
glueballs on their large ensemble of unquenched (Wilson gauge and quark action) 
configurations. They find 
little change in the masses of these states as the sea quark mass runs down
from the strange quark mass to near where the $\rho$ meson becomes unstable 
($m_\pi/m_\rho \approx 0.5$) but they do notice enhanced finite-volume
dependence which they attribute to dynamical quark effects. 

\section{HEAVY-QUARK HYBRID MESONS \label{sec:hybrid}}
  If heavy exotic mesons are to be seen as experimental resonances, the 
masses relative to threshold for decays into two heavy-light mesons are 
important in determining their widths. Constituent glue/quark models
predict \cite{Tani83} that the dominant decay mode of the lightest exotic, 
$1^{-+}$ will be into a pair of mesons, where one of the mesons has $L=1$ 
orbital angular momentum and so, in the model, if the hybrid is above this 
level it will probably be too broad to be seen. For the $\bbg$ hybrid, this 
channel has a 
threshold at 10.98 GeV (1.52 GeV above the $\Upsilon$ ground state). 
Nevertheless, the reliability of this model is questionable and if the hybrid 
is above the 2$B$ meson threshold it may still be a broad state. 

\subsection{Excited gluonic potentials}
  The most straightforward way of studying heavy mesons is by use of
the Born-Oppenheimer approximation. Since the fast-moving gluonic degrees of 
freedom have much shorter relaxation times than the slow-moving heavy quarks, 
the spectrum of this system can be approximated by solving Schr\"odinger's 
equation with the static potential as the interaction term. Hybrid mesons can 
be constructed by finding similar solutions with the inter-connecting
parallel transport now transforming 
under some non-trivial representation of the symmetry group of the 
static-source pair. These representations include excitations with a non-zero 
component of spin along the $Q\bar{Q}$ axis, for example. 

  Anisotropic lattice simulations have been used to study excitations to 
the
potential \cite{Juge97}. These modes are created on the
lattice by connecting the static quark propagators (temporal Wilson lines) with
linear combinations of path-ordered products of links that transform
irreducibly and non-trivially under the ``little group'' of lattice rotations 
about the $Q\bar{Q}$ molecular axis, combined with two discrete symmetries (for
details, see {\it e.g.} Ref. \cite{Juge97}). 
The standard ground state potential has quantum numbers $\Sigma_g^+$ and
the lowest excitation is the
$\Pi_u$, which carries one unit of angular momentum about the
$Q\bar{Q}$ axis. In the static approximation, where the quark spins do not
interact dynamically with the gluons,
a hybrid meson of two quarks bound in this interaction
potential can have a range of quantum numbers which include the exotic 
$0^{+-}$ and $1^{-+}$.
The new data are in agreement with existing Wilson action
results \cite{Pera90} and a number of previously unresolved 
potentials have also been computed. A larger range of inter-quark 
separations, up to 2.5 fm, has been studied.

Using empirical fits to the Monte-Carlo data for the $\Sigma_g^+$
and $\Pi_u$ potentials in the radial Schr\"odinger equation, the mass
of the lightest $b$-quark hybrid is 1.3 GeV above the $\Upsilon$(1S).
The leading Born-Oppenheimer approximation neglects retardation
effects and, above the two-B-meson energy where the lightest hybrid lies, the 
agreement with the experimental $b\bar{b}$ spectrum is poorer than 
below. The wave function of the potential model is found to be much larger than 
the small $\Upsilon$ meson. The peak in the radial probability density occurs 
at $\approx 0.6$ fm for the wave function in the $\Pi_u$ potential, compared 
with
$\approx 0.2$ fm for the ground state solution.  This suggests that the 
lattice volume for hybrid studies needs to be larger than for conventional 
mesons. It also implies that a good choice of operator to create a ground 
state hybrid would be a highly extended object. 

  The rich spectrum of excited potentials has been investigated \cite{Kuti97} 
within a diaelectric (bag-like) model of the QCD vacuum, giving results showing 
broad agreement with the Monte-Carlo data. In this model, the excitation 
energies are estimated by solving for the spectrum of confined gluon (and 
multi-gluon) states. This confinement is within a region where the physical 
vacuum has been expelled by the strong chromoelectric fields of the static-quark
pair. 

\subsection{Hybrid mesons from NRQCD}

\begin{figure}[t]
\leavevmode
\epsfxsize=2.9in\epsfbox{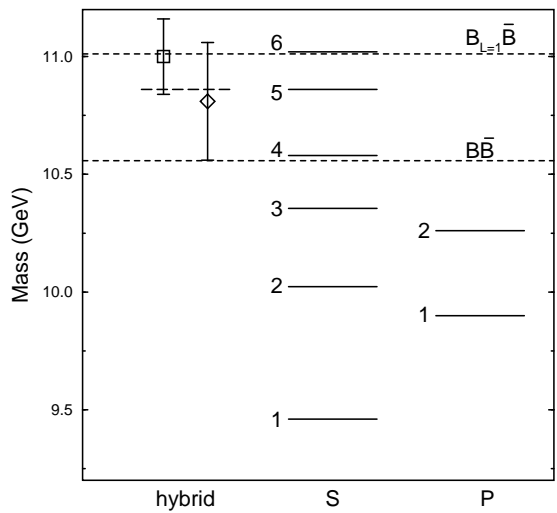}
\caption{Quenched lattice predictions for the lightest exotic $b\bar{b}g$ 
($1^{-+}$) hybrid. 
Solid lines are experimental data for S- and P-wave conventional mesons. The
two thresholds are indicated with dashed lines.
NRQCD data are from Refs. \protect{\cite{Mank97}} ($\Box$) and 
\protect{\cite{Davi97}} ($\diamond$) and the long-dashed line is the
potential model prediction \protect{\cite{Juge97}}\label{fig:bbar}}
\end{figure}

  Two groups \cite{Mank97,Davi97} presented results to the conference from 
NRQCD simulations of $b\bar{b}g$ hybrid mesons which are included in Fig.
\ref{fig:bbar}. These simulations, both performed on $16^3 \times 48$ lattices 
at $\beta = 6.0$, find the $1^{-+}$ hybrid excitation to be 1.64(16) GeV and 
1.35(25) GeV respectively above the 
ground state, suggesting the hybrid lies between the two thresholds
noted in Sec \ref{sec:hybrid}. Both these groups used large,
smeared operators.
These results are also included in Fig.~\ref{fig:bbar} and are 
consistent with the data from the Born-Oppenheimer approximation. 
In Ref. \cite{Davi97} simulations were performed with a higher-order-NRQCD 
action, and investigated the spin-dependent effects; preliminary data for the
$0^{+-}$ exotic follow model predictions that it is heavier than the 
$1^{-+}$. 

  NRQCD simulations on anisotropic lattices are in progress and the results of 
some exploratory work were presented to the conference \cite{Juge97}. 

\section{CONCLUSIONS}
  Recent developments in the use of anisotropic lattices have extended the
computational advantages of coarse lattices to studies of glueballs and 
heavy-quark hybrids. An anisotropic lattice, with a fine temporal spacing, 
allows 
Euclidean correlators to be resolved clearly over a range of time slices, while 
preserving most of the economic advantages of coarse lattices by keeping the 
spatial lattice spacing large. The problems with the scalar glueball on a 
coarse lattice are still not fully resolved, however some first results from 
new actions suggest a solution might be found while retaining the convenience
of the Symanzik improvement programme. Work on NRQCD and relativistic quark 
actions \cite{Alfo97} on
anisotropic lattices should extend these benefits to hybrid studies from the 
static approximation down to the light-quark sector.

  The preliminary findings from the higher states in the glueball spectrum
suggest a large number of glueball states at energies in the charmonium 
spectrum range. Details will be reported elsewhere. Other properties of 
glueballs, such as the matrix elements relevant for their creation in radiative 
charmonium decay, are now being studied on anisotropic lattices.

  The first results from NRQCD actions of $\bbg$ hybrids have shown that
these systems can be simulated reliably. While the data still have large
uncertainties, including the prospect of large finite-volume corrections due to
the highly extended nature of these objects, they suggest the heavy hybrid lies 
between the two thresholds (creation of 2 S-wave B mesons and creation 
of an S-wave and P-wave B meson pair). 

\section{ACKNOWLEDGEMENTS}

I am indebted to my collaborator, Colin Morningstar. I thank G.~Bali, 
S.~Collins, K.J.~Juge, T.~Manke and H.~Trottier for trusting me to present
their data and T.~Draper, J.~Kuti, K.F.~Liu, C.~Michael, and 
J.~Sloan for many helpful discussions. I am grateful to UKCCS for financial 
support. 



\end{document}